\documentclass[twoside,twocolumn,9pt]{article}
\usepackage{extsizes}
\usepackage[super,sort&compress,comma]{natbib} 
\usepackage[version=3]{mhchem}
\usepackage[left=1.5cm, right=1.5cm, top=1.785cm, bottom=2.0cm]{geometry}
\usepackage{balance}
\usepackage{widetext}
\usepackage{times,mathptmx}
\usepackage{sectsty}
\usepackage{graphicx} 
\usepackage{lastpage}
\usepackage[format=plain,ju stification=raggedright,singlelinecheck=false,font={stretch=1.125,small,sf},labelfont=bf,labelsep=space]{caption}
\usepackage{float}
\usepackage{fancyhdr}
\usepackage{fnpos}
\usepackage[english]{babel}
\usepackage{array}
\usepackage{droidsans}
\usepackage{charter}
\usepackage[T1]{fontenc}
\usepackage[usenames,dvipsnames]{xcolor}
\usepackage{setspace}
\usepackage{siunitx}
\DeclareSIPrefix\micro{\ensuremath{\upmu}}{-6}
\usepackage[compact]{titlesec}
\usepackage{bm,amsmath,amssymb,xfrac,float,url,soul}
\usepackage[flushleft]{threeparttable}
\usepackage[colorlinks=true,linkcolor=blue]{hyperref}
\definecolor{cream}{RGB}{222,217,201}
\usepackage{makecell}
\usepackage{commath}
\usepackage[normalem]{ulem}
\usepackage{amssymb}
\usepackage{soul}
\floatstyle{plaintop}
\restylefloat{table}
\usepackage{chemformula}
\usepackage{wasysym}
\usepackage{verbatim}
\usepackage{xcolor}
\usepackage{upgreek}

\graphicspath{{figures/}}

\begin{document}
\pagestyle{fancy}
\thispagestyle{plain}
\fancypagestyle{plain}{
\renewcommand{\headrulewidth}{0pt}}
\makeFNbottom
\makeatletter
\renewcommand\LARGE{\@setfontsize\LARGE{15pt}{17}}
\renewcommand\Large{\@setfontsize\Large{12pt}{14}}
\renewcommand\large{\@setfontsize\large{10pt}{12}}
\renewcommand\footnotesize{\@setfontsize\footnotesize{7pt}{10}}
\makeatother
\renewcommand{\thefootnote}{\fnsymbol{footnote}}
\renewcommand\footnoterule{\vspace*{1pt} 
\color{cream}\hrule width 3.5in height 0.4pt \color{black}\vspace*{5pt}} 
\setcounter{secnumdepth}{5}
\makeatletter 
\renewcommand\@biblabel[1]{#1}            
\renewcommand\@makefntext[1]
{\noindent\makebox[0pt][r]{\@thefnmark\,}#1}
\makeatother 
\renewcommand{\figurename}{\small{Figure}~}
\sectionfont{\sffamily\Large}
\subsectionfont{\normalsize}
\subsubsectionfont{\bf}
\setstretch{1.125}
\setlength{\skip\footins}{0.8cm}
\setlength{\footnotesep}{0.25cm}
\setlength{\jot}{10pt}
\titlespacing*{\section}{0pt}{4pt}{4pt}
\titlespacing*{\subsection}{0pt}{15pt}{1pt}
\fancyfoot{}
\fancyfoot[RO]{\footnotesize{\sffamily{1--\pageref{LastPage} ~\textbar  \hspace{2pt}\thepage}}}
\fancyfoot[LE]{\footnotesize{\sffamily{\thepage~\textbar\hspace{2pt} 1--\pageref{LastPage}}}}
\fancyhead{}
\renewcommand{\headrulewidth}{0pt} 
\renewcommand{\footrulewidth}{0pt}
\setlength{\arrayrulewidth}{1pt}
\setlength{\columnsep}{6.5mm}
\setlength\bibsep{1pt}
\makeatletter 
\newlength{\figrulesep} 
\setlength{\figrulesep}{0.5\textfloatsep} 
\newcommand{\topfigrule}{\vspace*{-1pt} 
\noindent{\color{cream}\rule[-\figrulesep]{\columnwidth}{1.5pt}} }
\newcommand{\botfigrule}{\vspace*{-2pt} 
\noindent{\color{cream}\rule[\figrulesep]{\columnwidth}{1.5pt}} }
\newcommand{\dblfigrule}{\vspace*{-1pt}
\noindent{\color{cream}\rule[-\figrulesep]{\textwidth}{1.5pt}} }
\makeatother

\def\an#1{\textcolor[rgb]{0.5,0,1}{#1}}
\def\td#1{\textcolor[rgb]{0.8,0.0,0.0}{#1}}

\twocolumn[
  \begin{@twocolumnfalse}
\par  
\sffamily

\noindent\LARGE{\textbf{Pisarenko's Formula for the Thermopower$^{\dag}$}}
\vspace{2mm}

\noindent\large{Andrei Novitskii,\textit{$^{a}$} Takao Mori\textit{$^{a,b,\ast}$}}
\vspace{2mm}
 
\noindent\normalsize{The thermopower $\alpha$ (also known as the Seebeck coefficient) is one of the most fundamental material characteristics for understanding charge carrier transport in thermoelectric materials. Here, we revisit the Pisarenko formula for the thermopower, which was traditionally considered valid only for non-degenerate semiconductors. We demonstrate that regardless of the dominating scattering mechanism, the Pisarenko formula describes accurately enough the relationship between thermopower $\alpha$ and charge carrier concentration $n$ beyond the non-degenerate limit. Moreover, the Pisarenko formula provides a simple thermopower-conductivity relation, $\alpha = \pm \frac{k_{\mathrm{B}}}{e} (b - \ln \sigma)$, where $b$ is a constant determined by the scattering mechanism and weighted mobility $\mu_w$, and $\sigma$ is the electrical conductivity. This relation is valid for materials with $\alpha > \SI{90}{\micro\volt\per\kelvin}$ when acoustic phonon scattering is predominant. This offers an alternative way to analyze electron transport when Hall measurements are difficult or inaccessible. Additionally, we show how the Pisarenko formula can be used to estimate the maximum power factor of a thermoelectric material from the weighted mobility of a single, not necessarily optimized, sample at any given temperature.
}
\vspace{2mm}

\end{@twocolumnfalse} \vspace{0.6cm}
]

\renewcommand*\rmdefault{bch}\normalfont\upshape
\rmfamily
\section*{}
\vspace{-1cm}
\footnotetext{$^{a}$Research Center for Materials Nanoarchitectonics (MANA), National Institute for Materials Science (NIMS), 1-1 Namiki, Tsukuba, Ibaraki, 305-0044, Japan.}
\footnotetext{$^{b}$Graduate School of Pure and Applied Sciences, University of Tsukuba, 1-1-1 Tennodai, Tsukuba, Ibaraki, 305-8573, Japan.}
\footnotetext{$^\ast$E-mail: 
MORI.Takao@nims.go.jp.}
\footnotetext{$^\dag$Electronic supplementary information available.}

\rmfamily


Thermoelectric materials are able to directly interconvert heat energy and electrical power, making them of particular interest for applications in solid-state cooling and power generation.\cite{mao2021,hendricks2022} 
Thermoelectric materials research typically aims to identify or develop materials with the best possible thermoelectric efficiency, defined by the figure of merit $zT = \alpha^2 \sigma T /\left(\kappa_{el}+\kappa_{lat}\right)$, where $\alpha$, $\sigma$, $T$ and $\kappa_{el}$ with $\kappa_{lat}$ represent the thermopower, electrical conductivity, temperature, electronic and lattice (phonon) thermal conductivities, respectively. 
Improving $zT$ is a non-trivial task due to the interrelationship among $\alpha$, $\sigma$, $\kappa_{lat}$, and $\kappa_{el}$, mainly correlated through the charge carrier concentration $n$ (chemical potential $\mu$) and the scattering mechanisms involved.\cite{ioffe1957,snyder2008} 
Therefore, understanding charge carrier transport properties is essential for engineering thermoelectric materials and enhancing their performance. Thermopower $\alpha$, in turn, is a fundamental material parameter particularly helpful to understanding and optimizing thermoelectric materials. $\alpha$ depends on the band structure, charge carrier transport entropy, scattering mechanism, and the doping level of the material, leading its quantity to be a collection of many characteristics, including band degeneracy, effective mass, scattering factor, and chemical potential.\cite{wang2023}

The effective mass model is often used to analyze experimentally measured thermopower. For many semiconductors, charge carrier transport can be adequately described by considering the states near the band edge.\cite{fistul1969,kireev1978,askerov1994} When parabolic dispersion is assumed, the thermopower can be considered as the measure of the chemical potential $\mu$ (refer to the Supporting Information file for details).\cite{takeuchi2010} Experimental thermopower measurements, in turn, are relatively straightforward to perform and highly widespread, particularly in laboratories focused on thermoelectric materials research. 
A common approach is to fit the experimental data to the form expected for a semiconductor with a single parabolic band, where charge carrier transport is dominated by a single scattering mechanism (e.g., acoustic phonon scattering), and the effective mass $m^{\ast}_d$ independent of doping level and temperature.
However, many materials deviate from this idealized scenario,\cite{deBoor2021} exhibiting multi-valley band structures,\cite{singh1997,may2009} non-parabolic bands,\cite{singh1994} complex scattering mechanisms,\cite{ravich1970} or disorder that affects the carrier transport.\cite{zvyagin1984,gantmakher2005} Nevertheless, even in such cases lying beyond the single parabolic band approximation, the effective mass model remains one of the most simple yet useful tools for transport properties analysis, guiding rational design toward better thermoelectric performance.\cite{zhang2020,wang2022}

Within the effective mass model, where carrier transport is dominated by majority carriers and described by a single parabolic band with a single scattering mechanism, the carrier relaxation time $\tau$ can be expressed by a simple power-law $\tau(\varepsilon)=\tau_{0}\varepsilon^{r}$ with $r$ representing the scattering factor. Consequently, the thermopower at any temperature or doping level can be described as a function of only chemical potential:\cite{fistul1969,askerov1994}

\begin{equation} \label{Eq:alpha_SPB}
\alpha\left(\eta\right)=\pm\frac{k_{\mathrm{B}}}{e}\left(\frac{\left(r+\sfrac{5}{2}\right)F_{r+\sfrac{3}{2}}\left(\eta\right)}{\left(r+\sfrac{3}{2}\right)F_{r+\sfrac{1}{2}}\left(\eta\right)}-\eta\right),
\end{equation}
where $k\mathrm{_B}$ is the Boltzmann constant, $e$ is the electron charge, and $F_{j}\left(\eta\right)$ is the $j$-th order Fermi integrals defined as
\begin{equation} \label{Eq:Fermi}
F_{j}\left(\eta\right) = \int_{0}^{\infty} \frac{\varepsilon^{j}}{1+e^{\varepsilon-\eta}} d\varepsilon,
\end{equation}
with $\varepsilon = E/k\mathrm{_B}T$ representing the reduced carrier energy, and $\eta = \mu/k\mathrm{_B}T$ representing the reduced chemical potential. The charge carrier concentration in this context is given by
\begin{equation} \label{Eq:conc_SPB}
n\left(\eta\right)=4\pi\left(\frac{2m^{\ast}_d k_{\mathrm{B}} T}{h^2}\right)^{\sfrac{3}{2}}F_{\sfrac{1}{2}}\left(\eta\right),
\end{equation}
where $m^{\ast}_d$ is the density-of-state effective mass, $h$ is the Planck constant.
Therefore, the effective mass $m^{\ast}_d$ value can be estimated from experimental values of thermopower $\alpha$ and carrier concentration $n$. For practical purposes, however, it is convenient to use analytical expressions that directly connect $\alpha$ and $n$, avoiding numerical integration of the Fermi integrals (Eq.~\ref{Eq:Fermi}). For the two limiting cases of a degenerate ($\eta \geq 5$) and a non-degenerate ($\eta \leq -1$) semiconductors, the Fermi integrals can be solved analytically, providing relatively simple expressions for the transport coefficients (for details, refer to Supporting Information; Figure~S1). Note that the band edge ($\eta = 0$) is used as the reference point for the chemical potential.\cite{kireev1978,fistul1969}

In the degenerate limit, Fermi integrals (Eq.~\ref{Eq:Fermi}) can be expressed as a power series through the Sommerfeld expansion, thereby reducing Eqs.~\ref{Eq:alpha_SPB} and \ref{Eq:conc_SPB} to
\begin{equation} \label{Eq:alpha_SPBdeg}
\alpha\left(\eta\right)=\pm\frac{k_{\mathrm{B}}}{e}\frac{\pi^{2}}{3}\frac{r+\sfrac{3}{2}}{\eta} \quad (\eta \geq 5)
\end{equation}
and
\begin{equation}\label{Eq:conc_SPB_deg}
n\left(\eta\right) = \frac{8\pi}{3} \left ( \frac{2m_d^{\ast}k_\text{B}T}{h^2} \right )^{3/2} \eta^{3/2} \quad (\eta \geq 5).
\end{equation}
By combining Eqs.~\ref{Eq:alpha_SPBdeg} and \ref{Eq:conc_SPB_deg}, one obtains the well-known formula for the thermopower in degenerate semiconductors:\cite{askerov1994}
\begin{equation}\label{Eq:alpha_deg}
\alpha = \pm\frac{8\pi^{2}k_{\text{B}}^{2}T}{3eh^2}m_{d}^{\ast}\left ( \frac{\pi}{3n} \right )^{2/3}\left ( r+\frac{3}{2} \right ).
\end{equation}

\begin{figure}[!t]
    \centering
    \includegraphics[width=1\linewidth]{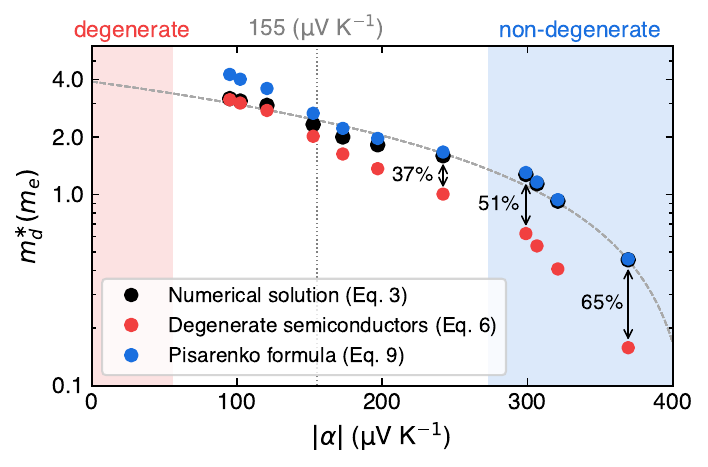}
    \caption{
    Effective mass $m^{\ast}_d$ as a function of the thermopower $\alpha$, calculated from the experimental data of \ch{Yb_{\textit{x}}Co4Sb12} (Ref.~\citenum{tang2015}) using the numerical solution (Eq.~\ref{Eq:conc_SPB}, black symbols), and two analytical formulas, namely, for degenerate semiconductors (Eq.~\ref{Eq:alpha_deg}, red symbols) and Pisarenko formula (Eq.~\ref{Eq:Pisarenko}, blue symbols). Acoustic phonon scattering was assumed ($r = -1/2$). The vertical gray dotted line indicates the threshold $\alpha$ value ($\approx \SI{155}{\micro\volt\per\kelvin}$), below which Eq.~\ref{Eq:alpha_deg} provides more accurate $m^{\ast}_d$ estimates. The black arrows indicate underestimation of $m^{\ast}_d$ when Eq.~\ref{Eq:alpha_deg} is used for $\alpha > \SI{155}{\micro\volt\per\kelvin}$. The gray dashed curve is a guide to the eye to visualize the general trend.}
    \label{Fig:Effective_mass_alpha}
\end{figure}

While some thermoelectric materials can indeed be considered degenerate semiconductors, a significant number of state-of-the-art systems are actually only partially degenerate (i.e., $-1 < \eta < 5$).\cite{ioffe1957} Partial degeneracy means that a semiconductor can exhibit features of both degenerate and non-degenerate regimes, e.g., metallic-like behavior of $\sigma(T)$ along with relatively high thermopower ($|\alpha| \geq \SI{150}{\micro\volt\per\kelvin}$). This combination, indeed, is frequently observed in well-known thermoelectric materials such as {\ch{Co4Sb12}}-based skutterudites,\cite{rogl2014,khan2017} {\ch{Mg3(Sb,Bi)2}} alloys,\cite{liu2021,imasato2022} \ch{BiCuSeO} oxyselenides,\cite{zhao2014} and others. Moreover, properly optimized materials usually have a chemical potential $\eta$ close to the band edge ($\eta_{opt} \approx 0$, $\alpha_{opt} \approx \SI{200}{\micro\volt\per\kelvin}$),\cite{zhang2020,kang2017a} which can result in significant underestimation of the effective mass by up to $\SI{30}{\percent}$ when using Eq.~\ref{Eq:alpha_deg} (Figures~\ref{Fig:Effective_mass_alpha}, S2).

In the non-degenerate region, Fermi-Dirac statistics can be replaced by Boltzmann statistics, simplifying Eq.~\ref{Eq:alpha_SPB} for the thermopower to
\begin{equation} \label{Eq:alpha_SPBnondeg}
\alpha\left(\eta\right)=\pm\frac{k_{\mathrm{B}}}{e}\left(r+\frac{5}{2}-\eta\right) \quad (\eta \leq -1).
\end{equation}
Similarly, the carrier concentration in the bandgap is expressed as 
\begin{equation} \label{Eq:conc_SPBnondeg}
n\left(\eta\right)=2\left(\frac{2\pi m^{\ast}_d k_{\mathrm{B}} T}{h^2}\right)^{\sfrac{3}{2}}e^{\eta} \quad (\eta \leq -1).
\end{equation}
Accordingly, using both Eqs.~\ref{Eq:alpha_SPBnondeg} and \ref{Eq:conc_SPBnondeg}, the relation between $\alpha$ and $n$ for non-degenerate semiconductors could be expressed as:
\begin{equation} \label{Eq:Pisarenko}
\alpha = \pm\frac{k_{\mathrm{B}}}{e}\left ( r + \frac{5}{2} + \mathrm{ln}\left[\frac{2\left ( 2\pi m^{\ast}_d k_{\mathrm{B}} T \right )^{3/2}}{h^3 n}\right] \right ).
\end{equation}
This relation, commonly referred to as the Pisarenko formula, is believed to have been derived by Nikolai Pisarenko between 1936 and 1940. However, the historical attribution of the Pisarenko formula (Eq.~\ref{Eq:Pisarenko}) remains unclear. While it is indeed widely associated with Pisarenko, there is no evidence that he ever published this formula.\cite{200yearsofTE2024} The first explicit mention of Pisarenko's contributions to thermoelectricity appears in the 1940 review paper by Davydov and Shmushkevich, where they noted that most of the formulas describing galvanomagnetic, thermomagnetic, and thermoelectric effects (including Eq.~\ref{Eq:Pisarenko}) in non-degenerate semiconductors were derived by Pisarenko.\cite{davydov1940}
However, Matvei Bronstein apparently first studied these effects in non-degenerate semiconductors in 1932 -- 1933.\cite{bronstein1932,bronstein1933} In his papers, Bronstein provided the classical relation for $\alpha(\eta)$ in the non-degenerate limit (Eq.~\ref{Eq:alpha_SPBnondeg}), as well as a precursor to the Pisarenko formula connecting $\alpha$ and $n$. Given the overlapping timeline, it is possible that Pisarenko and Bronstein collaborated or worked in parallel,\cite{frenkel1985} but the historical records remain ambiguous. It is worth noting that Bronstein's contributions to semiconductor physics were cut short by his execution in 1938 during the Great Purge, and citing his works during that period may have been considered dangerous.\cite{gorelik1994} Even so, Davydov and Shmushkevich included references to Bronstein's research in their 1940 paper, indicating its foundational importance. The question of Pisarenko's specific role in deriving the relation that now bears his name thus remains an open topic for historical investigation.\cite{Burkov2022}

\begin{figure}[!b]
    \centering
    \includegraphics[width=1\linewidth]{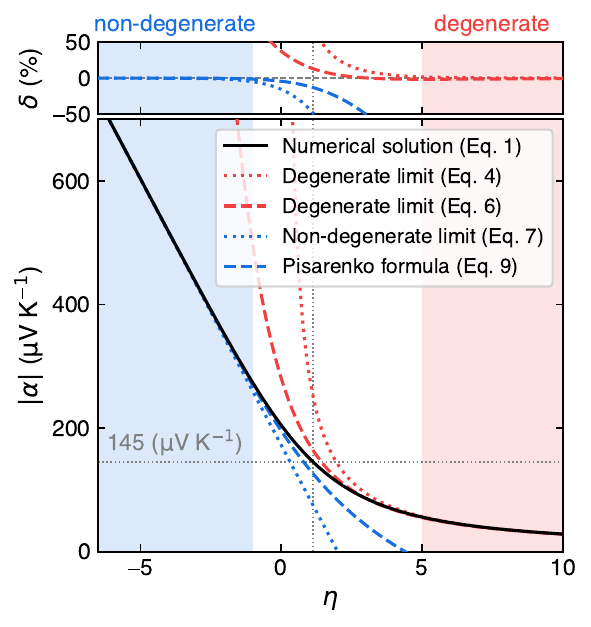}
    \caption{\label{Fig:Thermopower_eta}
    Thermopower $\alpha$ as a function of the chemical potential $\eta$, calculated using the numerical solution (Eq.~\ref{Eq:alpha_SPB}, black solid curve), the degenerate limit (Eq.~\ref{Eq:alpha_SPBdeg}, red dotted curve), the non-degenerate limit (Eq.~\ref{Eq:alpha_SPBnondeg}, blue dotted curve), and two analytical formulas for degenerate (Eq.~\ref{Eq:alpha_deg}, red dashed curve) and non-degenerate (Pisarenko formula, Eq.~\ref{Eq:Pisarenko}, blue dashed curve) semiconductors within the acoustic phonon scattering approximation ($r = -1/2$).
    The upper panel shows $\delta$, the relative difference in $\alpha$ calculated using the corresponding formulas. The thin gray dotted lines indicate the threshold where $\delta$ between $\alpha$ values calculated from Eq.\ref{Eq:alpha_deg} and Eq.\ref{Eq:Pisarenko} is equal and reaches $\approx\SI{13}{\percent}$, representing the limits of their applicability.
    }
\end{figure}
\begin{figure}[!t]
    \centering
    \includegraphics[width=1\linewidth]{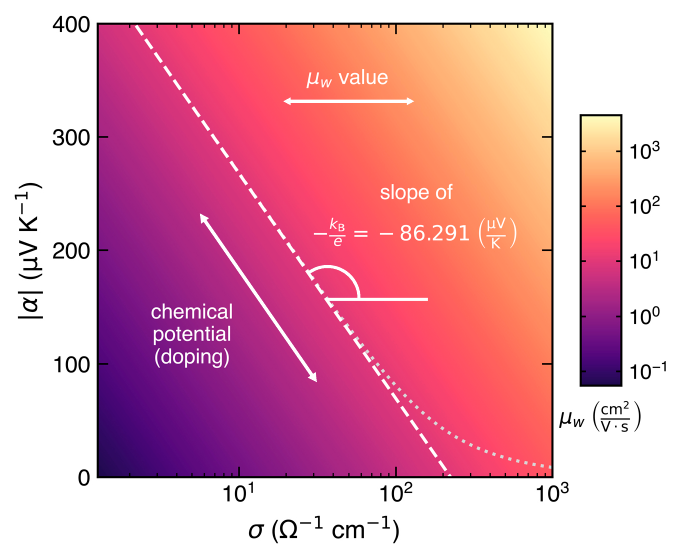}
    \caption{\label{Fig:Thermopower_sigma}
    A Jonker plot ($\alpha$ vs. log-scale $\sigma$), schematically displaying the thermopower-conductivity relation as predicted by the Pisarenko formula (Eq.~\ref{Eq:Pisarenko_sigma}, white dashed line) and the full numerical solution (Eq.~\ref{Eq:alpha_SPB}, light gray dotted curve). The slope of the Pisarenko line is $\pm k_{\mathrm{B}}/e$ $\approx$ $\pm \SI{86.291}{\micro\volt\per\kelvin}$ (positive slope for $n$-type semiconductors and a negative slope for $p$-type semiconductors), while its horizontal position is determined by the weighted mobility $\mu_w$ value.
    }
\end{figure}

At first glance, the limits of applicability of the Pisarenko formula (Eq.~\ref{Eq:Pisarenko}) should match those of the classical Bronstein formula (Eq.~\ref{Eq:alpha_SPBnondeg}). However, as initially noted by Yu.P.~Maslakovets and later confirmed by T.A.~Kontorova,\cite{kontorova1954} this is not the case. The Pisarenko formula (Eq.~\ref{Eq:Pisarenko}), in fact, agrees with the numerical solution (Eq.~\ref{Eq:alpha_SPB}) beyond non-degenerate limit up to $\eta \approx 1$ with an error of less than $\SI{10}{\percent}$ (Figure~\ref{Fig:Thermopower_eta}). 
As shown in Figure~\ref{Fig:Thermopower_eta} for the case of acoustic phonon scattering ($r = -1/2$), the partially degenerate regime spans a broad interval of thermopower values, ranging approximately from $\SI{56}{\micro\volt\per\kelvin}$ ($\eta = 5$) to $\SI{272}{\micro\volt\per\kelvin}$ ($\eta = -1$). Within this range, both analytical formulas, Eq.~\ref{Eq:alpha_deg} and Eq.~\ref{Eq:Pisarenko}, exhibit reasonable agreement with the full numerical solution from the corresponding sides, each maintaining a relative deviation below $\SI{10}{\percent}$ up to $|\alpha| \approx \SI{145}{\micro\volt\per\kelvin}$ (Figure~\ref{Fig:Thermopower_eta}).
Therefore, the Pisarenko formula accurately describes $\alpha (n)$ relationship for materials with $|\alpha| > \SI{150}{\micro\volt\per\kelvin}$ (Figure~S3). 
For higher degrees of degeneracy ($\eta > 1$ and $|\alpha| < \SI{150}{\micro\volt\per\kelvin}$), Eq.~\ref{Eq:alpha_deg} for degenerate semiconductors provides a better accuracy (Figures~\ref{Fig:Thermopower_eta}, S3).
This result is independent of temperature or the value of the effective mass. The only parameter affecting the agreement between the analytical formulas and the full numerical solution in the intermediate degeneracy region is the scattering factor $r$.
Nevertheless, since the qualitative trends in $\alpha(\eta)$ and $\alpha(n)$ remain similar regardless of the scattering mechanism, we focused on presenting and discussing results for acoustic phonon scattering ($r = -1/2$) in the main text, as it is still the most commonly observed dominant scattering mechanism in thermoelectric semiconductors. Yet, considering a growing body of studies suggesting polar optical phonon ($r = 1/2$) or ionized impurity ($r = 3/2$) scattering as dominant mechanisms in certain classes of thermoelectric materials,\cite{wang2015,shuai2017,shi2019,ren2020,ganose2022} we also provide results for $r = 1/2$ and $r = 3/2$ in the Supporting Information. We demonstrate that the Pisarenko formula (Eq.~\ref{Eq:Pisarenko}) remains valid for $\alpha$ values above $130-\SI{150}{\micro\volt\per\kelvin}$ regardless of the scattering mechanism (Figures~S4, S5). 
That said, it is important to emphasize that for thermoelectric materials with $|\alpha| > \SI{150}{\micro\volt\per\kelvin}$, the effective mass $m^{\ast}_d$ should be estimated using the Pisarenko formula (Eq.~\ref{Eq:Pisarenko}) rather than the more commonly used expression for degenerate semiconductors (Eq.~\ref{Eq:alpha_deg}), to avoid potentially significant underestimation, as demonstrated in Figure~\ref{Fig:Effective_mass_alpha}.

However, analyzing the thermopower $\alpha$ as a function of carrier concentration $n$ is limited since it still requires data from Hall measurements. For materials with high electrical resistivity and/or low charge carrier mobility, Hall measurements can be difficult or even impossible to perform. Additionally, the interpretation of the Hall data can be complicated for materials with a pronounced contribution from the anomalous Hall effect. 
From a practical perspective, it is much more convenient to analyze the relationship between $\alpha$ and $\sigma$, both of which can be readily measured over a wide temperature range, whereas Hall measurements are typically performed at or below room temperature. Similar to how $\alpha(n)$ relations are derived (Eqs.~\ref{Eq:alpha_deg} and \ref{Eq:Pisarenko}), $\alpha(\sigma)$ relations can also be obtained for both degenerate and non-degenerate regions (see Supporting Information for details):
\begin{equation}\label{Eq:alpha_deg_sigma}
\alpha=\pm\frac{k_{\mathrm{B}}}{e}\frac{\pi^{2}}{3}\left( r+\frac{3}{2} \right)\left( \frac{\sigma_{E_0}}{\sigma} \right)^{\frac{1}{r+3/2}}
\end{equation}
and
\begin{equation} \label{Eq:Pisarenko_sigma}
    \alpha = \pm\frac{k_\text{B}}{e} \left(r+\frac{5}{2}+\ln\left[{\Gamma\left( r + \frac{5}{2} \right)\frac{\sigma_{E_0}}{\sigma}}\right] \right),
\end{equation}
respectively. Here, $\Gamma$ is the gamma function, and $\sigma_{E_0}$ is the transport coefficient, which is merely a function of the weighted mobility $\mu_w$:
\begin{equation} \label{Eq:transport_coefficient}
    \sigma_{E_0} = \frac{8\pi e\left (2 m_{e}k_\text{B}T\right )^{3/2}}{3h^3}\mu_w.
\end{equation}
The weighted mobility is given by $\mu_w = \mu_{0}\left(m_{d}^{\ast}/m_e\right)^{3/2}$, where $\mu_0$ represents the intrinsic carrier mobility, and $m_e$ is the electron mass. Essentially, weighted mobility can be thought of as a descriptor of a material's inherent electronic transport properties, determining the maximum achievable power factor at a given temperature, as will be shown later. Furthermore, the weighted mobility can be considered an experimental parameter that, unlike the effective mass, can be easily calculated from the measured thermopower and electrical conductivity.\cite{snyder2020}

\begin{figure}[!t]
    \centering
    \includegraphics[width=1\linewidth]{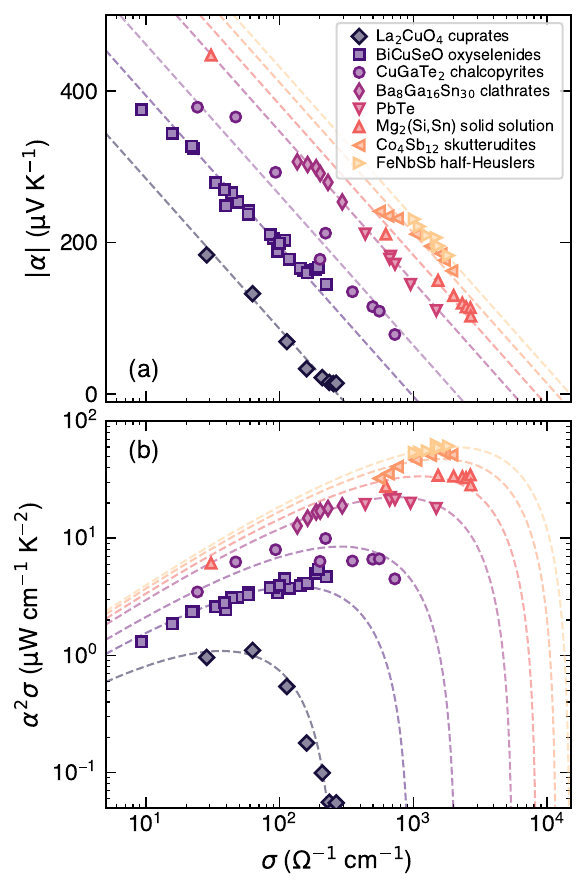}
    \caption{\label{Fig:Pisarenko_PF}
    Experimental (a) absolute thermopower $|\alpha|$ and (b) power factor $\alpha^{2}\sigma$ as functions of electrical conductivity $\sigma$ for selected thermoelectric material systems at various temperatures, including \ch{La2CuO4} (\SI{1123}{\kelvin}),\cite{su1990} \ch{BiCuSeO} (\SI{800}{\kelvin}),\cite{zhao2010,liu2015,feng2020a} \ch{CuGaTe2} (\SI{475}{\kelvin}),\cite{shen2016,ahmed2017} \ch{Ba8Ga16Sn30} (\SI{500}{\kelvin}),\cite{saiga2012} \ch{PbTe} (\SI{700}{\kelvin}),\cite{pei2014} \ch{Mg2(Si,Sn)} (\SI{300}{\kelvin}),\cite{liu2014} \ch{Co4Sb12} (\SI{800}{\kelvin}),\cite{shi2011,tang2014} and \ch{FeNbSb} (\SI{800}{\kelvin}).\cite{fu2015}
    Dashed lines represent Pisarenko formula based predictions (Eqs.~\ref{Eq:Pisarenko_sigma} and \ref{Eq:PF_simple}, respectively), calculated using different values of weighted mobility $\mu_w$ providing the best agreement with each experimental dataset.
    A portion of the literature data was retrieved from the StarryData2 database.\cite{katsura2019data,katsura2025}
    }
\end{figure}

Eq.~\ref{Eq:Pisarenko_sigma} provides a simple form of the thermopower-conductivity relationship:
\begin{equation} \label{Eq:Pisarenko_simple}
    \alpha = \pm \frac{k_{\mathrm{B}}}{e} (b - \ln \sigma)
\end{equation}
with
\begin{equation} \label{Eq:b}
    b = r+\frac{5}{2}+\ln\left[{\Gamma\left( r + \frac{5}{2} \right)\sigma_{E_0}}\right]
\end{equation}
and can be represented as a straight line with a slope of $\pm k_{\mathrm{B}}/e \approx \pm \SI{86.291}{\micro\volt\per\kelvin}$ (positive slope for $n$-type semiconductors and a negative slope for $p$-type semiconductors) on a linear $\alpha$ vs. log $\sigma$ scale (Jonker plot, Figure~\ref{Fig:Thermopower_sigma}).
In the case of acoustic phonon scattering, the $\alpha(\sigma)$ relationship provided by the Pisarenko formula (Eq.~\ref{Eq:Pisarenko_sigma}) agrees with the numerical solution (Eq.~\ref{Eq:alpha_SPB}) down to $|\alpha| \approx \SI{90}{\micro\volt\per\kelvin}$ (Figures~\ref{Fig:Thermopower_sigma}, S6). It should be noted, however, that the agreement is notably less accurate for other scattering mechanisms. For polar optical phonon scattering ($r = 1/2$) and ionized impurity scattering ($r = 3/2$), the lower bounds for reliable applicability of Eq.~\ref{Eq:Pisarenko_sigma} increase to \SI{168}{\micro\volt\per\kelvin} and \SI{236}{\micro\volt\per\kelvin}, respectively (Figure~S7).
Despite its virtual simplicity, the prediction given by the Pisarenko formula (Eq.~\ref{Eq:Pisarenko_sigma}), in general, is well substantiated by experiments, even though the materials under study were not necessarily single band or non-degenerate semiconductors (Figure~\ref{Fig:Pisarenko_PF}).\cite{jonker1968,su1990,rowe1995,kang2017}
In many such materials, indeed, the simultaneous change of $\alpha$ with $\sigma$ upon rigid-band-like doping follows a slow logarithmic decrease as shown in Figure~\ref{Fig:Pisarenko_PF}a.\cite{ohtaki1996,toberer2008,toberer2010,zhu2011,ahmed2017,guelou2020,khanina2024} 
Moreover, as noted earlier, there exists a significant group of partially degenerate semiconductors in which the charge carrier concentration $n$, as well as the effective mass $m^{\ast}_d$, remain constant or change only slightly with temperature. For these materials, the Pisarenko formula can reasonably capture not only $\alpha(n)$ and $\alpha(\sigma)$ trends, but also the temperature dependence of the thermopower (Figure~S8).
Furthermore, considering the temperature dependence of the chemical potential,\cite{takeuchi2010} or including the contribution from minority carriers,\cite{johnson1953,lautz1953} allows the Pisarenko formula to describe the temperature dependence of the thermopower in a broad range of semiconducting materials with reasonable accuracy.\cite{collignon2020,nakajima2024}
In this context, the Jonker-type analysis can serve not only as an alternative to Hall measurements when they are not accessible, but rather as a complement that may offer additional insights.

\begin{figure}[!t]
    \centering
    \includegraphics[width=1\linewidth]{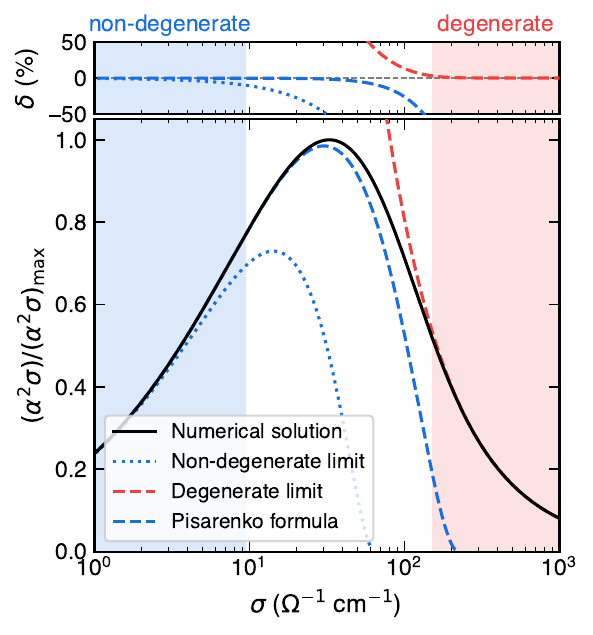}
    \caption{\label{Fig:PF_sigma}
    Normalized power factor $(\alpha^{2}\sigma)/(\alpha^{2}\sigma)_{\mathrm{max}}$ as a function of electrical conductivity $\sigma$ calculated using the full numerical solution (black solid curve), the non-degenerate limit (blue dotted curve), the degenerate limit (red dashed curve), and the Pisarenko formula (Eq.~\ref{Eq:PF_simple}, blue dashed curve) within the acoustic phonon scattering approximation ($r = -1/2$). The upper panel shows $\delta$, the relative difference in $(\alpha^{2}\sigma)/(\alpha^{2}\sigma)_{\mathrm{max}}$ calculated using the corresponding formulas.
    }
\end{figure}

In case of acoustic phonon scattering ($r=-1/2$), the applicability range of the Pisarenko formula (Eq.~\ref{Eq:Pisarenko_sigma}) should be also sufficient to accurately describe the power factor $\alpha^2 \sigma$ and predict its maximum, expected at $\eta \approx 0.67$ within the effective mass model. Considering Eq.~\ref{Eq:Pisarenko_simple}, the power factor can be expressed as 
\begin{equation} \label{Eq:PF_simple}
    \alpha^2 \sigma = \left(\frac{k_{\mathrm{B}}}{e}\right)^2 \sigma (b^2 - 2b \ln\sigma + (\ln\sigma)^2).
\end{equation}
This expression, derived from the Pisarenko formula, agrees well with the full numerical solution and underestimates the maximum value of the power factor by only about $\SI{1.5}{\percent}$ (Figure~\ref{Fig:PF_sigma}). The maximum power factor, in turn, can be determined from the condition $\frac{\partial (\alpha^2 \sigma)}{\partial\sigma} = 0$, which yields $\ln\sigma = b - 2$. Therefore, in this simplified theoretical model, $(\alpha^2 \sigma)_{\mathrm{max}}$ depends only on $r$ and $\mu_w$ (or $\sigma_{E_0}$):
\begin{equation} \label{Eq:PF_max}
    (\alpha^2 \sigma)_{\mathrm{max}} = 4\left(\frac{k_{\mathrm{B}}}{e}\right)^2 e^{r+1/2} \Gamma \left(r+\frac{5}{2}\right)\sigma_{E_0}.
\end{equation}
Considering $r=-1/2$, Eq.~\ref{Eq:PF_max} can be rewritten as
\begin{equation} \label{Eq:PF_max_mu_w_full}
    (\alpha^2 \sigma)_{\mathrm{max}} = 4\left(\frac{k_{\mathrm{B}}}{e}\right)^2 \frac{8\pi e\left (2 m_{e}k_\text{B}T\right )^{3/2}}{3h^3}\mu_w
\end{equation}
or
\begin{equation} \label{Eq:PF_max_mu_w_short}
    (\alpha^2 \sigma)_{\mathrm{max}} = 1.73\cdot 10^{-5} T^{3/2}\mu_w,
\end{equation}
where $(\alpha^2 \sigma)_{\mathrm{max}}$ is in $\SI{}{\micro\watt\per\cm\per\square\kelvin}$ and $\mu_w$ in $\SI{}{\square\cm\per\volt\per\s}$.
Hence, given only the electrical conductivity and thermopower of a single sample at an arbitrary doping level, Eq.~\ref{Eq:PF_max_mu_w_short} can provide a reasonable estimate of its highest power factor expected after optimizing its carrier concentration (Figure~\ref{Fig:Pisarenko_PF}b).

Considering the aforementioned and the increasing ease, availability, and quality of electrical transport measurements, namely, $\alpha$ and $\sigma$, this type of analysis can become a preferred first step in the characterization of many thermoelectric materials. The amount of information that can be extracted from just one measurement is substantial. In this context, the Pisarenko formula provides a simple yet robust analytical link between thermopower and carrier concentration (or conductivity) under the effective mass approximation. 
While traditionally associated with non-degenerate semiconductors, we have shown that this formula remains reasonably accurate well beyond the non-degenerate limit. In particular, for materials with $|\alpha| > \SI{150}{\micro\volt\per\kelvin}$, the Pisarenko formula (Eq.~\ref{Eq:Pisarenko}) yields a more accurate estimate of the effective mass than the widely used expression for degenerate semiconductors (Eq.~\ref{Eq:alpha_deg}). Likewise, the thermopower-conductivity relation derived from the Pisarenko formula (Eq.~\ref{Eq:Pisarenko_sigma}) remains applicable down to $|\alpha| \approx \SI{90}{\micro\volt\per\kelvin}$ for acoustic phonon scattering, and enables practical analysis of charge carrier transport even in the absence of Hall measurements (Figure~\ref{Fig:Pisarenko_PF}).
Even in the era of first-principles calculations and advanced Boltzmann transport modeling, the Pisarenko formula retains conceptual value as it can offer a first-order description of a material's electronic structure, reveal trends, or serve as a baseline for identifying anomalous transport behavior.\cite{ahmed2017,serhiienko2024}
We emphasize that the goal of this paper is not to claim the universality of the Pisarenko formula, but rather to highlight its practical range of validity, practical limitations, and provide a coherent framework for its use in rational analysis of thermoelectric transport. We believe that such analysis remains a useful and accessible method for initial interpretation of the transport data, assessing a material’s optimization potential, and guiding more detailed computational investigations.\cite{ogata2019,settipalli2020,yamamoto2022,matsubara2025,akhtar2025}

\section*{Acknowledgements}

This work was supported by JST Mirai JPMJMI19A1. The authors are grateful to Prof.~Alexander Burkov (Ioffe Institute) and Prof.~Jean-Fran\c{c}ois Halet (University of Rennes) for fruitful discussions. A.N. also thanks Jenny Murakoshi, current and former Thermal Energy Materials Group (NIMS) members for plentiful discussions.

\section*{CRediT Statement}

\textbf{Andrei Novitskii}: Conceptualization, Methodology, Investigation, Data Curation, Writing (Original Draft), Writing (Review \& Editing), Visualization.
\textbf{Takao Mori}: Writing (Review \& Editing), Supervision, Funding Acquisition.

\section*{Conflicts of Interest}

There are no conflicts to declare.
\section*{Data Availability Statement}

Data sharing is not applicable to this article as no new data were created in this study.
\section*{Keywords}

Pisarenko, Thermopower, Seebeck Coefficient, Effective Mass Model, Thermoelectrics

\bibliography{refs}
\bibliographystyle{rsc}

\end{document}